\documentclass[nonacm,sigplan]{acmart}

\usepackage{algorithm}
\usepackage[noend]{algpseudocode}
\usepackage{graphicx}
\usepackage{textcomp}
\usepackage{xcolor}
\usepackage{pifont}
\usepackage{epsfig, wrapfig}
\usepackage{listings}
\usepackage{soul}
\usepackage{multirow}
\usepackage{makecell}
\usepackage{bm}
\usepackage[most]{tcolorbox}
\usepackage{tikz,tcolorbox}

\newcommand{\crmark}{\textcolor{green}{\ding{51}}} % Checkmark
\newcommand{\cwmark}{\textcolor{red}{\ding{51}}} % Checkmark
\newcommand{\xrmark}{\textcolor{green}{\ding{55}}} % Cross
\newcommand{\xwmark}{\textcolor{red}{\ding{55}}} % Cross
\begin{document}

\newcommand{\wfig}[5]{%
	\begin{wrapfigure}{#3}{#4\columnwidth}%
		\vspace{-3ex}%
		\begin{center}%
			\epsfig{file=#2, width = #4\columnwidth}%
		\end{center}%
		\vspace{-4ex}%
		\caption{#5.\label{figure:#1}}%
		\vspace{-2ex}%
\end{wrapfigure}}

\title{Towards Enabling Distance-Based Memory Addressing}

\author{Ananth Krishna Prasad}
\affiliation{%
  \institution{University of Utah}
  \city{Salt Lake City}
  \state{Utah}
  \country{USA}
}
\email{ananth@cs.utah.edu}

\author{Sharad Bhat}
\affiliation{%
  \institution{University of Utah}
  \city{Salt Lake City}
  \state{Utah}
  \country{USA}
}
\email{sharad.bhat@utah.edu}

\author{Rajeev Balasubramonian}
\affiliation{%
  \institution{University of Utah}
  \city{Salt Lake City}
  \state{Utah}
  \country{USA}
}
\email{rajeev@cs.utah.edu}

\author{Mahdi Nazm Bojnordi}
\affiliation{%
  \institution{University of Utah}
  \city{Salt Lake City}
  \state{Utah}
  \country{USA}
}
\email{mnbojnordi@gmail.com}

\begin{abstract}
Approximate Nearest-Neighbor Search (ANNS) in high dimensional vector datasets is an application of significant prevalence across different AI applications. However, such an operation is significantly bandwidth limited at large working-set sizes owing to the \textit{curse of dimensionality}. Traditional indices used to accelerate ANNS rely on search-space pruning as a preprocessing step to alleviate such bandwidth requirement, but such optimization occurs either at the cost of increased bandwidth-inefficiency and/or degradation of search quality. This paper proposes a data-parallel hardware/software mechanism for performing large-scale similarity search in-memory. We propose a novel algorithm to simplify the computation requirement for similarity search across various distance metrics through lightweight primitives to perform a fast and approximate data-parallel brute-force search on the entire vector space. We further build a memory system capable of executing the required operations to generate a distance metric per datapoints, which is then used to enable pruning as a post-processing step. We offer adequate software support for user control over the proposed system. By enabling such search-space pruning as a post-processing step, we achieve near-perfect recall across representative workloads while achieving orders of magnitude performance and energy improvement over state-of-the-art algorithmic approaches on million and billion-scale workloads.
\end{abstract}
\maketitle % should come after the abstract
\pagestyle{plain} % should come right after \maketitle

\section{Introduction}
\label{sec:introduction}

Nearest Neighbor Search (NN-Search) in multidimensional data space is a fundamental problem in computer science. The brute-force approach for finding the nearest neighbor for a given query involves calculating the distance to every point in the dataset (linear scan). Such an approach is memory-bound owing to the requirement of fetching the entire dataspace for each query.  Unfortunately, it is often impossible to provably find the exact nearest neighbors without resorting to brute force, especially as the dimensionality of data increases. This problem is termed popularly as \textit{the curse of dimensionality}. 
It turns out that most of the present-day applications of nearest neighbors can tolerate a certain amount of imprecision, leading to the emergence of Approximate Nearest Neighbor Search (ANNS). Multimedia such as text, audio, and video can be represented as embedding vectors indexed into a vector database. ANNS in such vector databases are prevalent in a variety of server-scale machine-learning applications such as Large-Language Models~\cite{NEURIPS2020_6b493230, mao2020generation} used in ChatGPT, information retrieval systems~\cite{xiong2020approximate, ludewig2018effective}, recommendation systems~\cite{suchal2010full, hasanzadeh2019improving}, and web search~\cite{vanderkam2013nearest}. ANNS algorithms generate indices on the dataset through pre-processing, which is leveraged to prune down the search space for each query.

State-of-the-art algorithms for high-dimensional ANNS can be categorized into two main classes - graph-based and compression-based. Compression-based indices use compression/quantization techniques to encode vectors using codebooks. This lowers the memory footprint and computational intensity per datapoint~\cite{FAISS, ANNA, matsui2018survey, babenko2014inverted}, but is bandwidth-bound and inaccurate. Graph-based algorithms~\cite{malkov2018efficient, matsui2018survey, ScaNN} build neighborhood graphs on top of the uncompressed vector dataset. This approach is highly accurate but not scalable due to significant indexing overhead and bandwidth inefficiency stemming from memory indirection during graph traversal. The root cause of inefficiency in the existing indices is in the inherently massive bandwidth required to overcome the curse of dimensionality. Irrespective of the underlying approach, it is often necessary to visit a large fraction of the vector dataset to ensure high recall, resulting in bandwidth-bound behavior.

This bandwidth-bound characteristic of ANNS applications presents a unique challenge to computer architects. One approach to addressing the bottleneck is to accelerate these indices using Processing-In-Memory and leverage the massive bandwidth available within memory devices. However, traditional indices incur memory indirection during pruning, making a poor fit for the SIMD-like computation model PIM offers. This, coupled with significant computation requirements renders conventional PIM inefficient in accelerating state-of-the-art ANNS indices.
To address the above problems, we propose a novel memory system for highly accurate similarity search applications through careful hardware-software codesign. We envision the proposed memory system to function as a "Distance-Addressable Memory", or DAM, where the memory is capable of calculating and returning the k-"closest" points to any input query point. The key idea is to utilize the massive internal bandwidth to perform a fast and approximate data-independent brute-force search within memory, then leverage the generated approximate distance metrics to prune the dataspace in a data-parallel fashion. By formulating pruning as a post-processing step to brute-force computation rather than as a pre-filtering step, we manage to mitigate the curse of dimensionality, offering fast similarity search within a high recall space. 

We make the following contributions: \textbf{1)} We propose a novel algorithm for similarity search that generates an approximate distance metric for vector datasets on a per-query basis.  With a simple reformulation of the distance calculation, we support multiple distance metrics. \textbf{2)} We propose a lightweight near-array processor (NAP) coupled with a 1T-1R-based memory array that computes approximate distances. These distances are then used as indices to generate partially ordered responses from the memory, enabling search-space pruning and bandwidth-efficient data reranking. \textbf{3)} We identify the challenges in the proposed system and devise data/control-flow techniques that improve efficiency. Adequate software support is provided through an API that enables sufficient user control over memory configuration. \textbf{4)} We highlight key metrics in emerging vector database applications, such as index-freshness, support for out-of-distribution queries, and filtered searches, and show how DAM readily extends support for such applications.

A simulation-based evaluation of our algorithm and platform over a wide variety of real-world embedding vector datasets shows that DAM achieves near-perfect accuracy with 52.7$\times$ and 97$\times$ throughput and latency improvement respectively over a state-of-the-art compression-based approach on an Intel Xeon E5-2680.
%on a variety of real-world billion-scale datasets. 
Each DAM chip consumes 50\% more area in comparison to a conventional RRAM chip, while achieving an energy improvement of $4.15\times$ on average.   
\section{Background and Motivation}

%This section covers necessary background information and motivates the problem.
%
\subsection{Similarity Search}

\paragraph{K-Nearest Neighbors} $k$-NN is a supervised learning technique used for the classification or regression of data. For a given dataset, the algorithm takes an input target point along with $k$ and returns the $k$ closest data points to the target point in terms of a defined distance metric.
\paragraph{The Curse of Dimensionality} The total volume of the dataspace increases exponentially with an increase in dimensionality. This results in very low-density high-dimensional datasets, rendering conventional data structures ineffective in accelerating exact K-NN. Therefore, the most efficient method for finding the exact $k$-nearest neighbors in a given $D$-dimensional data space (D $\geq$ 100) is through brute force search~\cite{yu2010high}, making it an $O(N\times D)$ operation per query. Given that both computation and bandwidth complexity are dependent on $N$, the brute-force approach is prohibitively expensive at large dataset sizes.
\paragraph{ANN Search} With a relaxed constraint on the exactness of the nearest neighbors, it is possible to index high-dimensional data to find top-$k$ "similar" data points to any given query. Approximate Nearest Neighbor Search (ANNS), or Similarity Search, is an umbrella term used for a range of mechanisms that query across a very large space of objects to find "similar" objects using a defined distance metric. A range of algorithms have been proposed to tackle the ANNS problem defined below: %A definition of the parameters that dictate the quality of any ANNS algorithm is explained below
Consider a query q with $d$ dimensions, and a dataset D with $d$ dimensional data points and suppose the algorithm outputs a set X of k candidate near neighbors, and G is the ground-truth set of the k closest neighbors to q. Then, we define the recall-$k@k$ of this set $\mathbf{X}$ to be $\frac{|X \cap G|}{k}$. The goal of an ANNS algorithm then is to maximize recall while retrieving the results as quickly as possible. There is an inherent trade-off between recall and latency/QPS in any ANN algorithm. Different algorithms operate in different regions within this trade-off space. 
Most rely on either a compression-based index or a graph-based index.

\subsubsection{Compression-Based Index} Compression-based indices, e.g., product quantization, rely on data encoding to compress the data space. They have two major benefits: 1) reduced memory footprint of the total working set and 2) simplified distance computation through reduced data bitwidth. %The most popular quantization method for compression-based indexing is product quantization.

\begin{figure}[h!]

	\begin{center}
		\epsfig{file=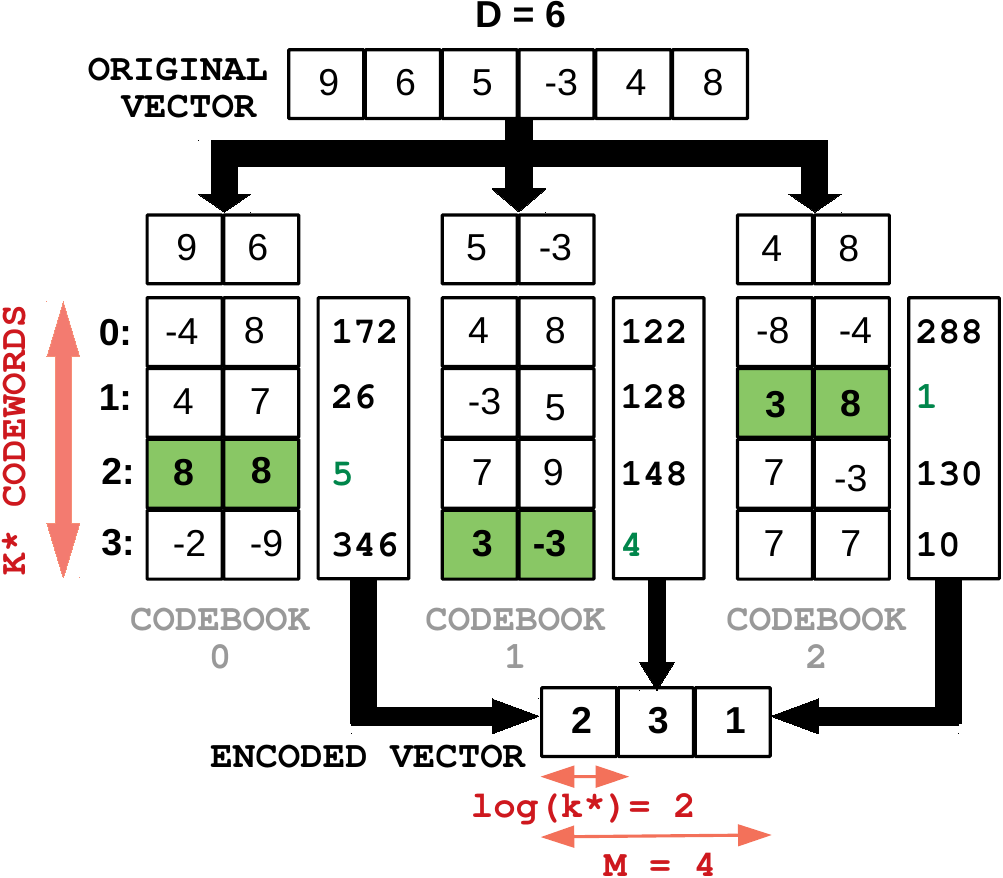, width = 0.6\columnwidth}
	\end{center}

	\caption{Illustrative example of product quantization
		\label{figure:PQ}}

\end{figure}

\paragraph{Product Quantization} Product Quantization (\textbf{PQ}) is a source coding technique to compress and encode vectors in a multi-dimensional vector space. The $D$-dimensional vector dataset is split into $M$ $D/M$ sub-dimensional vector subsets, and $k*$ centroids (or codewords) are extracted for each of the $M$ sub-dimensional subsets through clustering. These $k*$ codewords within each subset form a codebook. Each subvector within a vector subset is now encoded as the ID of the closest codeword within its corresponding codebook. Combination of all such encodings across all subsets gives us the final compressed representation of each vector. Each datapoint in the compressed representation requires $M\log k*$ bits. An example of such encoding is shown in Figure~\ref{figure:PQ} using $k*$ = $4$ and $M$ = $3$ on a $D$=$6$ dimensional vector.

Search over a raw product quantized dataset requires a brute-force approach, resulting in a high bandwidth requirement. State-of-the-art methods %for compression-based similarity search 
use an Inverted File Index (IVF) in combination with product quantization to mitigate bandwidth demands. IVF algorithms apply preprocessing to partition the entire vector space to generate $|C|$ clusters and their corresponding centroid set $C$. Product quantization is applied to each of the vector datasets within these clusters individually to compress the vectorspace.
%Each generated codeword across all $|C|$ clusters is encoded as a residue of its corresponding centroid

\paragraph{Search} The IVF helps in pruning down the search space by allowing an initial level of filtering through a selection of only the $nprobe$-closest centroids to each query. $nprobe$ is a search parameter that dictates the depth of the search. The search occurs over the $nprobe$ closest quantized spaces through a brute-force approach for each query, while maintaining a priority list for the top-$k$ candidates.

\subsubsection{Graph-Based Index}
Graph-based indices usually rely on some enhanced version of neighborhood small world (NSW) graphs. NSW graphs are formed by connecting each data point to all its nearest neighbors within a defined radius. Similarity search for a given query point involves greedy traversal across the graph. The accuracy of the search is a function of the depth of the search. High-recall operations require higher search depth, which leads to high bandwidth requirements and inefficiency. Different optimizations have been proposed to the NSW graph approach to alleviate the effect of high search depth, but such benefits are reaped at the cost of increased indexing overhead. The state-of-the-art approach for in-memory graph-based ANNS indexing is the hierarchical navigable small-world graph (HNSW), which defines hierarchies within the NSW graph to allow for long links.

%However, HNSW suffered from scalability issues yet again owing to indexed graphs exceeding main-memory sizes. Every node in the graph has to store metadata such as its $N_{out}$ and level IDs to which it belongs, along with the datapoint it represents. For billion-scale datasets spanning hundreds of GBs in size, it becomes highly impractical to store such an index in memory.
%DiskANN~\cite{subramanya2019diskann} solves the above problem by constructing a graph with a low diameter to enable scaling-out of index onto SSDs. Further optimizations such as PQ are used to store a compressed form of the index within main memory to decrease SSD roundtrips. 

\subsection{Memristive Array Structure}
\wfig{rram}{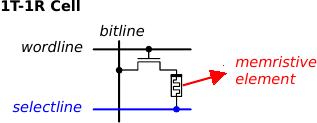}{r}{.4}{1T-1R memory cell~\cite{zangeneh2014design}}
Memristive technology has been promoted as an alternative to conventional memories due to their scalability, non-volatility, and being free of leakage power.
In particular, resistive RAM (RRAM) is one of the most promising memristive devices under commercial development that shows great potential for building main memory systems. Moreover, they have shown unique capabilities for efficient in-memory processing. Specifically, RRAM-based memory arrays are capable of back-end-of-line (BEOL) integration, where the RRAM cell material can be deposited on just a couple of metal layers to create a memory array, with the remaining layers available for traditional CMOS logic design. Numerous cell architectures have been proposed in the literature that optimize RRAM for better reliability, density, and computational capabilities.
1R crosspoints are denser but lack isolated access to individual rows and columns~\cite{xu2015overcoming}. As the proposed in-situ approach requires isolated column access a 1T1R memory cell (Figure~\ref{figure:rram}) is preferred over the 1R crosspoint.

\subsection{Motivation}
The working set size corresponds to the overall memory footprint of the data + index. PQ-based approaches have high bandwidth efficiency but are bandwidth-bound. Because they compress data to simplify the bandwidth-bound, they suffer from low recall-$k@k$ ($\le$ 0.6) on their performant configurations. This can mainly be attributed to two key reasons: 1) The centroid selection prunes out a significant fraction of the data points automatically on which the distance would never be computed. This is a common pitfall associated with the curse of dimensionality and 2) There is no access to the underlying raw data for re-ranking. Graph-based approaches offer the capability for high-recall through a more efficient pruning approach and support for re-ranking. However, Graph-based approaches impose significant indexing overhead on the underlying raw data, with each datapoint node requiring information regarding its connected neighbors. The search space pruning, which requires multiple-levels of graph traversal, is highly bandwidth-inefficient owing to multiple memory indirections. 
\begin{table}[h]
    \centering
    \begin{tabular}{|c|c|c|c|}
        \hline
        \textbf{Feature} & \textbf{Compression} & \textbf{Graph} & \textbf{Ours} \\ \hline
        Working Set Size & $\frac{1}{4}$ to $\frac{1}{16}$x & $2$ - $10$x~\cite{hnswfais67:online} & 1x\\ \hline
        Scalability & \crmark & \xwmark & \crmark\\ \hline
        Bandwidth Efficient & \crmark & \xwmark & \crmark \\ \hline
        Bandwidth-Bound & \cwmark & \cwmark & \xrmark \\ \hline
        High-Recall Support & \xwmark & \crmark & \crmark \\ \hline
        \multicolumn{4}{|c|}{\textbf{Emerging Applications}} \\ \hline
        OOD queries & \xwmark & \xwmark & \crmark \\ \hline
        Index Freshness & \xwmark & \xwmark & \crmark \\ \hline
        Filtered queries & \xwmark & \xwmark & \crmark \\ \hline
    \end{tabular}
    \caption{Compatibility of existing indexing approaches with the proposed approach for vector similarity search }
    \label{tab:features}

\end{table}
Moreover, current indices fail to generalize for emerging applications listed in table~\ref{tab:features}. Existing indices fail to generate high-recall for queries sampled out-of-distribution~\cite{jaiswal2022ooddiskann} (OOD) wrt the dataset. This is due to the data-dependent nature of index generation by existing algorithms. We hypothesize that a query-based pruning approach, enabled through pruning as a post-processing step, could generalize better for queries OOD. Index freshness~\cite{singh2021freshdiskann} refers to the capability of the index to incorporate new vectors within its domain. Existing approaches for the same rely on periodic index updates to account for fresh vectors, which incurs significant recall and throughput penalty. A query-based indexing approach offers support for per-query index generation for fresh vectors. Filtered queries~\cite{singh2021freshdiskann} requires indices to search only across a subset of the original dimensions. Because conventional indices leverage pairwise distances across vectors to extract neighborhood information, such an index does not capture behavior under just a subset of dimensions.
Table~\ref{tab:features} summarizes our motivation.

\section{Distance-Aware Pruning}

% \begin{figure*}[t]
% 	\begin{center}
% 		\epsfig{file=figures/Algorithm.pdf, width = 0.8\linewidth}
% 	\end{center}
%
% 	\caption{
% (a) Proposed runtime quantization depending on bitmask $M$ (b) Illustrative examples of candidate selection with the proposed search algorithm in 1D (c) Aggregation of per-dimension distance to generate final distance metric
% 		\label{figure:1DBin}}
% \end{figure*}

In short, state-of-the-art algorithms suffer from the curse of dimensionality owing to one common reason: they perform an initial pre-filtering step and perform subsequent distance computation ONLY on the filtered dataset. To avoid the same pitfall, we take an alternative approach that performs fast and approximate brute force computation on the entire dataspace, then performs filtering as a post-processing step. By leveraging the large internal bandwidth to perform both the distance computation and dataspace pruning, we avoid being bandwidth-bound.
\
\begin{tcolorbox}[colback=red!5!white,colframe=red!75!black,title=Summary of Proposed Algorithm]
\small
This section describes the proposed algorithm in detail. We first leverage runtime-quantization through bitmasking to quantize the dataset by masking out the $\mathbf{M}$ least significant bits within each dimension. We then propose a simple data-parallel distance computation mechanism that relies on just 3 basic operations: comparison/XNOR, counting, and addition. We then show how any distance metric can be calculated in an approximate fashion with only these three operations through a simple reformulation of distance calculation. Once these distance values per-vector are generated, we show how to leverage them to prune the dataspace in a data-parallel fashion.
\end{tcolorbox}

%\textcolor{red}{The above paragraph is an excellent overview. Good candidate for putting into a colored box so it stands out.}

%As explained previously, one common feature across the different data-indexing methods for similarity search is to perform search-space pruning while performing a vector-wise search only on a small subset of the original data. If such pruning could be performed in-memory, the bandwidth inefficiency would be eliminated. 
% PQ-like approaches are great at light-weight indexing but operates at a low recall space due to lack of support for re-ranking. HNSW-like approaches can generate high-recall through re-ranking, but are limited by memory-indirection and the resulting bandwidth inefficiency.
%either through $nprobe$ in PQ and the $efsearch$ parameter HNSW 2)
%For the in-memory implementation, we would ideally like to combine the benefits of both approaches, with lightweight indexing on top of the raw data with support for re-ranking. Moreover, we need to enable indexing the vectors within each memory device in an independent fashion across the different memory devices to entirely eliminate memory-indirection. With these motivations in-mind, we design a memory-friendly algorithm for similarity search.
%We then devise an algorithm to calculate the distance across all quantized vectors given any query $\mathbf{Q}$ quantized in a similar fashion. We then demonstrate how the proposed algorithm can be executed in a data-parallel fashion through a generalized reformulation of the distance calculation.

\subsection{Runtime Quantization of Data}
\label{subsection:Quantization}

Consider a vector dataset $\mathbf{V}$ with $\mathbf{D}$-dimensional vectors and a $\mathbf{D}$-dimensional query $\mathbf{\tau}$, where each $ d \in  D \in [0, 2^N]$, with $\mathbf{N}$ being the representation bitwidth.

Runtime quantization of $\mathbf{V}$ relies on the intrinsic properties of fixed-point data. A bitmask applied to the least $\mathbf{M}$ bits of $\mathbf{V}$ quantizes the data from a $\mathbf{2^N}$ to a $\mathbf{2^{(N - M)}}$ representation space ($\mathbf{V^{\prime}}$), with each value representing $\mathbf{2^{M}}$ values from the $\mathbf{2^N}$ space. Such quantization can be used to compress the representation space of both signed and unsigned fixed-point numbers.

\begin{figure}[t]
	\begin{center}
		\epsfig{file=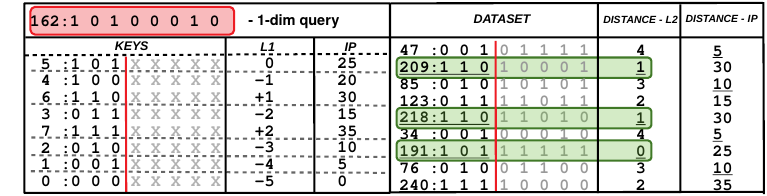, width = \linewidth}
	\end{center}

	\caption{Example Key Generation for a 1-dimensional query with search on an example dataset. The green highlighted vectors are identified as candidate point using Algorithm~\ref{algorithm:1dp}.
		\label{figure:1DBin}}

\end{figure}

\subsection{1-D Pruning Algorithm}

Our pruning method relies on simple operations on quantized data to extract the defined distance metric across each dimension. We first show how this works in a single dimension, and then extend it for multi-dimensional data.

Consider a target point $\tau = 162$. Consider an example query that requires us to find 3 similar data points to $\tau$ out of an example dataset $\mathbf{V}$ (shown in Figure~\ref{figure:1DBin}). After generating $\tau{\prime}$ and $V{\prime}$, we first identify every possible distance value in the quantized space and map it with the point that would generate the corresponding distance value in relation to $\tau{\prime}$. These points form our keys $K$, illustrated in Figure~\ref{figure:1DBin} with their corresponding L1 and IP distances. We compare each vector in $V^{\prime}$ with all generated keys $k \in K$ using bitwise XNOR operations in an iterative fashion, Once a key matches completely with a vector ($k = v^{\prime}$), the distance associated with $k$ is mapped to the vector. After iterating over all $K$, 3 similar vectors are given by the three vectors closest to the query in the quantized space. 
\begin{algorithm}
  \scriptsize
\caption{Initialize($N$, $V$, $\tau$, $M$) - Initialization\label{algorithm:1dinit}}
	\begin{algorithmic}[1]
            \State $\mathbf{I}\leftarrow [0, (2^{N - M} - 1)]$ \Comment{L1/I2}
            \State $\mathbf{I}\leftarrow [0, (2^{N - M} - 1)]$ \Comment{IP}
            \State $\mathbf{K}\leftarrow\{$Keys$\}$
		\State $\mathbf{Dist_{K}}\leftarrow\{$Distance per Key$\}$
		\State $\tau^{\prime}\leftarrow\tau / 2^{M}$, $\mathbf{V}\prime\leftarrow\mathbf{V} / 2^{M}$
            \State $\mathbf{Out_{V}}\leftarrow\{$Output distance per vector$ v \in \mathbf{V}\}$
        \algstore{1D-Prune-1}
        \end{algorithmic}
\end{algorithm}

% \begin{algorithm}
%   \scriptsize
% \caption{KeyGeneration($I$, $\tau^{\prime}$)\label{algorithm:1dc} - Key Generation\label{algorithm:1dkeygen}}
%     \begin{algorithmic}[1]
%     \algrestore{1D-Prune-1}
%     \ForAll{$i\in I$} 
%         \State $\mathbf{k}_{i}$ = $|i - \tau^{\prime}|$ \Comment{L-norm}
%         \State $Dist_{k} = i$
%         \If{$i \% \tau^{\prime} == 0$} \Comment{IP}
%             \State $\mathbf{k}_{i}$ = $|\frac{i}{\tau^{\prime}}|$
%             \State $Dist_{k} = i$
%         \EndIf
%     \EndFor
%     \algstore{1D-Prune-2}
%     \end{algorithmic}
% \end{algorithm}
\begin{algorithm}
  \scriptsize
\caption{KeyGeneration($I$, $\tau^{\prime}$, $D$, Metric)\label{algorithm:1dc} - Key Generation\label{algorithm:1dkeygen}}
    \begin{algorithmic}
    \algrestore{1D-Prune-1}
    \ForAll{$i\in I$}
        \ForAll{$d\in D$}
            \If{Metric = L2 \textbf{or} Metric = L1}
                \If{$\tau^{\prime}_{d} + i \le 2^{N-M}-1$}
                    \State $\mathbf{k}^{i,+}_{d} = \tau^{\prime}_{d} + i$ \Comment{Positive symmetric coordinate exists}
                \EndIf
                \If{$\tau^{\prime}_{d} - i \ge 0$}
                    \State $\mathbf{k}^{i,-}_{d} = \tau^{\prime}_{d} - i$ \Comment{Negative symmetric coordinate exists}
                \EndIf
                \If{Metric = L2}
                    \State $Dist_{k} = i^{2}$ \Comment{L2 squared contribution}
                \Else
                    \State $Dist_{k} = i$ \Comment{L1 contribution}
                \EndIf
            \ElsIf{Metric = IP \textbf{and} $i \% \tau^{\prime}_{d} == 0$} \Comment{IP}
                \State $\mathbf{k}^{i}_{d} = \frac{i}{\tau^{\prime}_{d}}$
                \State $Dist_{k} = i$
            \EndIf
        \EndFor
    \EndFor
    \algstore{1D-Prune-2}
    \end{algorithmic}
\end{algorithm}
% \begin{algorithm}
%   \scriptsize
% \caption{KeyGeneration($I$, $\tau^{\prime}$, $D$, Metric)\label{algorithm:1dc} - Key Generation\label{algorithm:1dkeygen}}
%     \begin{algorithmic}[1]
%     \algrestore{1D-Prune-1}
%     \ForAll{$i\in I$}
%         \ForAll{$d\in D$}
%             \If{Metric = L2 \textbf{or} Metric = L1}
%                 \State $\delta = \sqrt{i}$ \text{ (for L2) or } $i$ \text{ (for L1)} \Comment{Calculate coordinate offset}
%                 \State $\mathbf{k}^{i,+}_{d} = \tau^{\prime}_{d} + \delta$ \Comment{Positive symmetric coordinate}
%                 \State $\mathbf{k}^{i,-}_{d} = \max(0, \tau^{\prime}_{d} - \delta)$ \Comment{Negative symmetric coordinate (clamped)}
%                 \State $Dist_{k} = i$
%             \ElsIf{Metric = IP \textbf{and} $i \% \tau^{\prime}_{d} == 0$} \Comment{IP}
%                 \State $\mathbf{k}^{i}_{d} = \frac{i}{\tau^{\prime}_{d}}$
%                 \State $Dist_{k} = i$
%             \EndIf
%         \EndFor
%     \EndFor
%     \algstore{1D-Prune-2}
%     \end{algorithmic}
% \end{algorithm}

\begin{algorithm}
  \scriptsize
\caption{DistCalc($V^{\prime}$, $K$) - Distance Generation\label{algorithm:1dcompute}}
    \begin{algorithmic}[1]
    \algrestore{1D-Prune-2}
    \ForAll{$v\prime\in\mathbf{V}\prime$}
        \ForAll{$k\in\mathbf{K}$}
            \If{$(v^{\prime} == k)$}
              \State $Out_{v} = Dist_{k}$
            \EndIf
        \EndFor
    \EndFor
    \State \textbf{return} $map(V, Out)$
    \end{algorithmic}
\end{algorithm}

Algorithms~\ref{algorithm:1dinit} to~\ref{algorithm:1dp} formalize the proposed algorithm. The complete sequence is split into 4 different steps - Initialization, Key generation, Distance calcuation and Candidate selection.
\begin{algorithm}
  \scriptsize
\caption{DistPrune($map(\mathbf{V},\mathbf{Dist}), Metric, k$) - Candidate Selection\label{algorithm:1dp}}
	\begin{algorithmic}[1]
            \State $\mathbf{Out}\leftarrow\varnothing$
            \State $Count\leftarrow 0$
        
            \While{$Count < k$ \textbf{and} $\mathbf{V} \neq \varnothing$}
        		\ForAll{$v\in\mathbf{V}$}
                    \If{Metric = L2 \& $\mathbf{dist_{v}} = 0$} \Comment{L-norm}
		                \State $\mathbf{Out \leftarrow Out \cup v}$ 
                        \State $\mathbf{V \leftarrow V \setminus v}$ 
                        \State $Count = Count + 1$
                    \ElsIf{Metric = IP \& $\mathbf{dist_{v}}$ = NUM\_MAX}  \Comment{IP}
		                \State $\mathbf{Out \leftarrow Out \cup v}$
                        \State $\mathbf{V \leftarrow V \setminus v}$
                        \State $Count = Count + 1$
                    \Else
                        \If{Metric = L2}
                            \State $dist_{v} = dist_{v} - 1$ 
                        \ElsIf{Metric = IP}
                            \State $dist_{v} = dist_{v} + 1$
                        \EndIf
                    \EndIf

                    \If{Count = $k$}
                        \State \textbf{Exit For}
                    \EndIf
                \EndFor
            \EndWhile
            \State \textbf{return} $\mathbf{Out}$
	\end{algorithmic}
\end{algorithm}

\subsection{Multidimensional Distance Generation}

Given a multi-dimensional vector dataset, the DistCalc function can be naively applied on a vector dataset in a dimension-wise fashion to generate the manhattan and IP distance metrics from the query across each dimension. These per-dimension values need to be aggregated across dimensions for each vector point. This can be represented as iteration of algorithms~\ref{algorithm:1dkeygen} and~\ref{algorithm:1dcompute} over all dimensions in the dataset. This conventional formulation of the approximate distance calculation using dimension-wise distances $I_{L1}^{d}$ and $I_{IP}^{d}$ for a given vector is denoted in the below equations:
%Such naive application of the proposed algorithm on the vector dataset would generate the distance on a per-dimension basis for each vector,
\begin{equation}
\label{equation: basicDist}
    I_{L2} = {\sum_{d=1}^{D} \left(I_{L1}^{d}\right)}^{2} /\\
    I_{IP} = {\sum_{d=1}^{D} I_{IP}^{d}} \\
\end{equation}

\begin{figure}[h]

	\begin{center}
		\epsfig{file=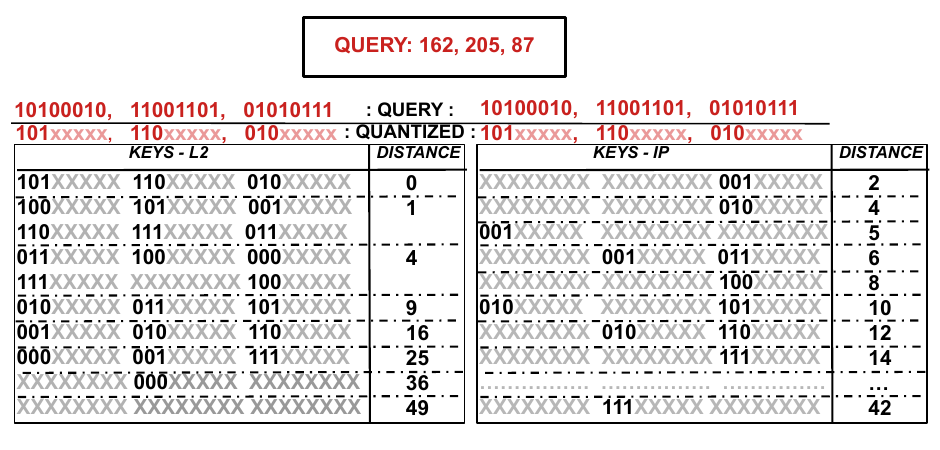, width = \linewidth}
	\end{center}

	\caption{An example showing key-ordering for search for multidimensional search for a given query with $N=8$ and $M=5$
		\label{figure:key_ordering}}

\end{figure}

However, such an approach is not SIMD-friendly owing to irregular tree-like access pattern required for aggregation. To facilitate easier adaptation of the proposed algorithm in a data-parallel fashion, we reformulate the distance calculation. In equations~\ref{equation: basicDist}, we observe that the range of values possible for $i \in I$ is bounded within $[0, (2^{N-M})^{2}]$ for both L2 and Inner Product. Using the above observation, it helps to rewrite the distance calculation equation as a sum of the number of all possible per-dimension distances multiplied by the count of each unique distance:
\begin{equation}
\label{equation: countbased}
    I_{Dist} = \sum_{k=0}^{K} c_{k} \times f(k)
\end{equation}
where $I_{Dist}$ represents the unified final quantized distance, $K$ represents a multi-dimensional key, where each dimension corresponds to the same distance $f(k)$ from the given query, and $c_{k}$ represents the count of the number of matching dimensions between the quantized vector and query for the corresponding key. This forms the key idea behind our hardware implementation of the proposed algorithms. Such an implementation requires minor tweaks to the DistCalc function, shown in the highlighted region in algorithm~\ref{algorithm:multidcompute}.

\begin{algorithm}
  \scriptsize
\caption{DistCalcMultiD($V^{\prime}$, $K$) - Distance Generation\label{algorithm:multidcompute}}
    \begin{algorithmic}[1]
    \ForAll{$v\prime\in\mathbf{V\prime}$}
        \State $Out_{v} \leftarrow 0$
        \ForAll{$k\in\mathbf{K}$}
        \State $c(k) \leftarrow 0$
        \ForAll{$d\in\mathbf{K_{D}}$}
            \If{$(v^{\prime}[d] == k[d])$}
              \State $c(k)$ += $1$
            \EndIf
        \EndFor
        \State $Out_{v}$   $+= c(k)\times Dist_{k}$
        \EndFor
    \EndFor
    \State \textbf{return} $map(V, Out)$
    \end{algorithmic}
\end{algorithm}

\paragraph{Search Key Generation and Ordering} Algorithm~\ref{algorithm:1dkeygen} generates the map between each key and it corresponding distance for each dimension. A union operation across all such maps gives us the unified map for the multi-dimensional query. Such an example is shown in Figure~\ref{figure:key_ordering} for a 3-dimensional datapoint. Though the upper limit on the total number of possible distances is given as $(2^{N-M})^{2}$, empirically, the number of distances per each query containing a valid key is significantly lower. Moreover, the number of iterations of line 3 in algorithm~\ref{algorithm:multidcompute} is a factor of the number of valid per-dimension values in a key rather than all dimensions, eliminating ineffectual comparisons.
%\textcolor{red}{This last paragraph is getting into some nuances. Here would be a good place to have a Summary box with Section 3 takeaways.}
\section{DAM Architecture}

% The proposed algorithm has three key desirable features that makes it a good fit for processing-in-memory:
% \begin{itemize}
%     \item The algorithm is capable of performing the computation in a data-parallel/SIMD fashion to leverage the massive internal bandwidth provided by the memory system.
%     \item The underlying computation to realize the algorithm is lightweight. The entire algorithm requires just 3 basic arithmetic/logic operations - XNOR/Comparison, Count, and Addition. Any fixed-point multiplication can be implemented through iterative additions to reduce complexity.
%     \item The algorithm incurs requires minimal bandwidth owing and incurs no memory indirection, thereby removing any bandwidth-dependent stalls.
% \end{itemize}

This section describes how the proposed similarity search algorithm can be implemented in memory through in-situ primitives coupled with lightweight additional logic. High-density memristive cells can be integrated with CMOS-logic using BEOL integration~\cite{9265083, 9088053}, which makes them amenable to low-cost peripheral circuits that are not easily compatible with DRAM cells. We first introduce a 1T-1R memristor-based memory array (DAM array) capable of performing the required in-situ operations, then extend the array with the proposed logic for scaling and accumulating dimension-wise distance metrics. 

\subsection{DAM Memory Array}

\begin{figure}[h]

	\begin{center}
		\epsfig{file=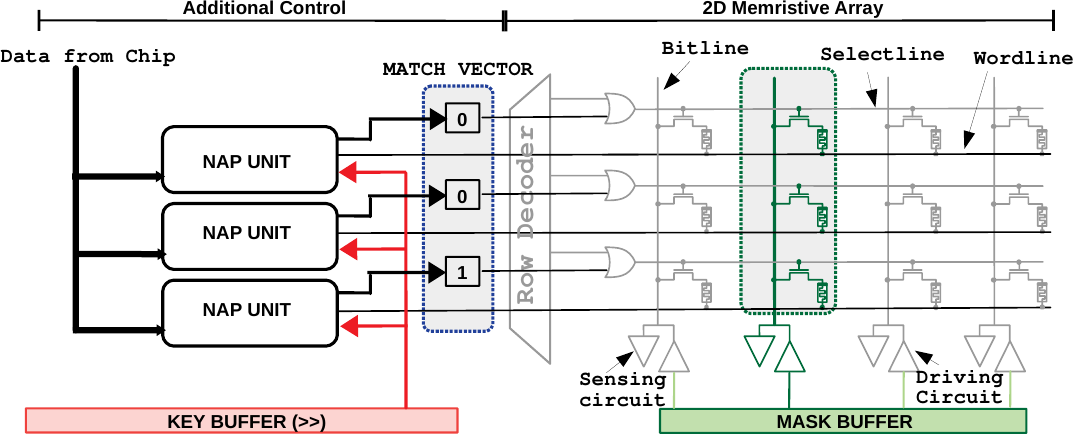, width = 0.9\linewidth}
	\end{center}

	\caption{DAM Memory Array
		\label{figure:array_overview}}

\end{figure}

Figure~\ref{figure:array_overview} shows the DAM memristive array capable of data-parallel bit-serial computation. The Sensing and Driving circuit connected to each memory bitline enables it to perform normal reads and writes. When a specific bitline is being driven, the bitline is charged high (shown in green in the figure). The selectlines are driven by the match vector, which indicate all rows containing the target dataset within the array.

\paragraph{Bitwise Column Read} For each selected row, any driven bitline gets discharged through the wordline if the corresponding memristive element is in a low-resistance ($R_{L}$) state. Similarly, the bitline does not get discharged if the row is in a high-resistance ($R_{H}$) state. Hence, at the end of every bitline search, the corresponding wordlines return a column-wise read of data across all selected rows. To create a near-ideal situation for bitwise read and avoid sneak currents, we choose memristive devices that provide a large dynamic range of resistance states (i.e., $R_{H}$ is much bigger than $R_{L}$). Additional logic is employed at the array periphery to generate and operate on the required column-wise reads to realize the proposed algorithm.

\paragraph{Additional Logic} The additional logic required to enable the proposed algorithm in-memory is illustrated in figure~\ref{figure:array_overview}. A mask buffer is interfaced at the sensing circuit periphery to offer control over bitline selection and ordering. The mask buffer contains a bitmap which is serially traversed to perform the necessary bitline activations. This helps realize the runtime quantization of data in-memory.  A Near-Array-Processor (NAP) is interfaced with each wordline in the memory array, which is fed by bit-serial column-wise reads in the memory array. Each NAP performs computation on a series of masked reads and outputs a single bit-value indicating the corresponding row's candidacy status to the match vector. This match vector denotes the candidacy status of each memory row.

\subsection{Near-Array Processor}
\begin{figure*}[h]

	\begin{center}
		\epsfig{file=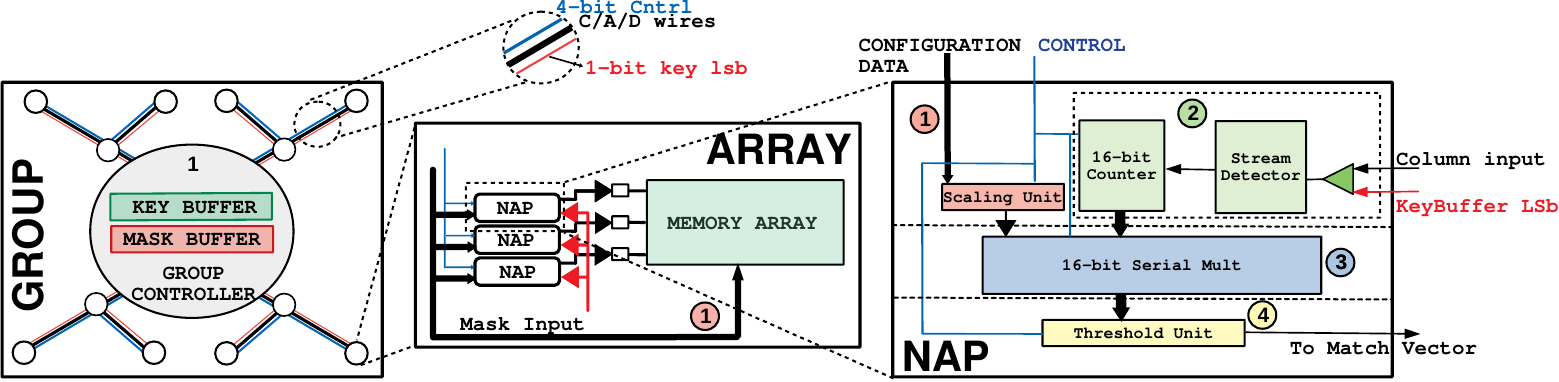, width = \linewidth}
	\end{center}

	\caption{Organization of a Memory Group. Each group consists of multiple arrays connected through a H-tree. The H-tree routs control, configuration parameters and mask to their corresponding destinations within each array and NAP. The different pipeline stages - denoted by shaded pattern per unit/buffer - are shown in the legend}
		\label{figure:mem_group}
\end{figure*}

The internals of each NAP unit are shown in Figure~\ref{figure:mem_group}. Each NAP unit contains an XNOR unit, a stream detection unit, an 8-bit counter, a 16-bit serial multiplier, and a 16-bit threshold counter on the input datapath from the memory arrays. Configuration parameters are multiplexed onto the memory write datapath from the chip to enable user control over the NAPs. The NAP output datapath updates the match vector to denote candidacy status. The purpose of each of the proposed hardware units is detailed in the paragraph below.

\textbf{Key and Mask Buffers:} The key and mask buffers (shown in Figure~\ref{figure:array_overview}) are 2048-bit wide registers. The Key Buffer, incorporated with right-shift capability, feeds its least significant bit to the XNOR units across all NAPs. Similarly, the mask buffer feeds the mask value to the array one byte at a time. The purpose of the mask buffer is to activate only the required bitlines within the memory array. After each masked bitread and comparison, the Key Buffer is right-shifted to update the LSb. \textbf{Stream Detection for dimension-wise key matching:}
The stream detector unit is designed to identify key match per dimension by analyzing every $\mathbf{N - M}$ consecutive output bits from the XNOR unit. If the $\mathbf{N - M}$ consecutive bits are all $1$, the stream detector outputs high and resets the pattern, denoting a key match within the corresponding dimension. \textbf{Count aggregation across dimensions per key vector:}
The Stream Detector unit feeds into a 16-bit counter after every $\mathbf{N - M}$ cycles, which keeps track of the count of key matches across all dimensions per key vector. This count represents $c(k)$ in equation~\ref{equation: countbased} after computation on the corresponding $k$ key-buffer entry. \textbf{Scaling up of distance through serial multiplication:}
The counter storing $c(k)$ for each $k$ feeds into a serial multiplication unit, which hosts a 16-bit buffered signed adder coupled with a scaling unit. The scaling unit contains a buffered counter that stores the value of $f(k)$ corresponding to each key. The adder is integrated with the scaling reg and counter to perform multiplication through a set of iterative additions. Once the computation is performed across all keys, the match vector is reset which initiates the Thresholding phase. At this point, the buffered adder contains $I_{dist}$, which is the approximate distance value based on the given distance metric. \textbf{Thresholding for candidate selection:} Once $I_{dist}$ is generated, the thresholding unit generates the final candidate set (Algorithm~\ref{algorithm:1dp}). The thresholding unit contains a buffered 16-bit counter. This counter can be configured to up-count or down-count, depending on the requirements of the distance metric and application as defined by the algorithm. On counter saturation, the corresponding match bit is set in the match vector, indicating candidacy. The threshold counters across all NAPs are reset once the required number of rows are read out, indicating end of search for the current query.

\subsection{Control-flow and Pipelining}
Multiple DAM arrays together form a memory group. A memory group is a single synchronized unit of compute sharing a single copy Key/Mask buffers across its constituent arrays and their NAPs. Each group employs a controller that communicates through the H-tree to orchestrate operations across all NAPs and arrays, and to facilitate candidate data transfer through automatic index generation. This section describes the group controller's design and operation.

\subsubsection{Control flow}
The group controller imposes a control-flow on its constituent NAPs and arrays based on the requirements of the proposed algorithm. Such control is responsible for orchestration of the different functional units across all NAPs and arrays. Each NAP is pipelined to ensure maximal throughput. The different pipeline stages are highlighted in Figure~\ref{figure:mem_group}. Sufficient double buffering is enabled at the periphery of each unit within the NAP to enable the required pipeline. We discuss each of the pipeline stages and how the computation is overlapped across the different stages.

\paragraph{Pipelining} Figure~\ref{figure:mem_group} highlights the different pipelening stages within a memory group. \textbf{Stage 1} corresponds to the transfer of new key/scale/bitmask and mask values into their corresponding buffers to initiate a search for a given query. The configuration parameters are broadcasted to all the NAPs under the group controller's domain, with each parameter being transferred in a serial fashion owing the the shared bus. Once these writes are performed, the chip controller initializes the bit-serial read across arrays, which initializes \textbf{Stage 2} of the pipeline. Once computation corresponding to the key/mask is completed and $c(k)$ is generated, the group controller instructs all NAPs to begin accumulation into the global counter, initializing \textbf{Stage 3} of the pipeline. Stage 3 performs accumulation of each per-scale count into the existing partial sum buffer, upon completion of which across all keys corresponding to a query writes into the threshold unit, which initializes \textbf{Stage 4} of the pipeline. During stage 4, the saturation of threshold unit counter results in a match vector update, which then signals data transfer availability for that specific row. Stage 4 is executed only once all keys corresponding to a query are done, and overlaps with the execution of keys corresponding to the next query.

\paragraph{Handling Stalls} The Key and Mask buffers at the group controller are shared across different arrays to keep the area overhead at a minimum. However, The central location of the group controller necessitates synchronous operation across all the arrays. Units done with a specific stage need to wait for the other units across all NAPs to complete their computation. The execution times of stages 1, 2 and 4 are the same across all arrays within the group. However, the execution time of the accumulation stage depends on $c(k)$, which might vary across arrays. To account for such pipeline stalls, all units present within each NAP send signals through the additional control bus denoting completion of the operation. This bit undergoes logical AND across all arrays within the group during upstream propagation, such that the group controller receives a completion signal only when all NAPs have completed the corresponding stage.

\begin{figure}[h]

	\begin{center}
		\epsfig{file=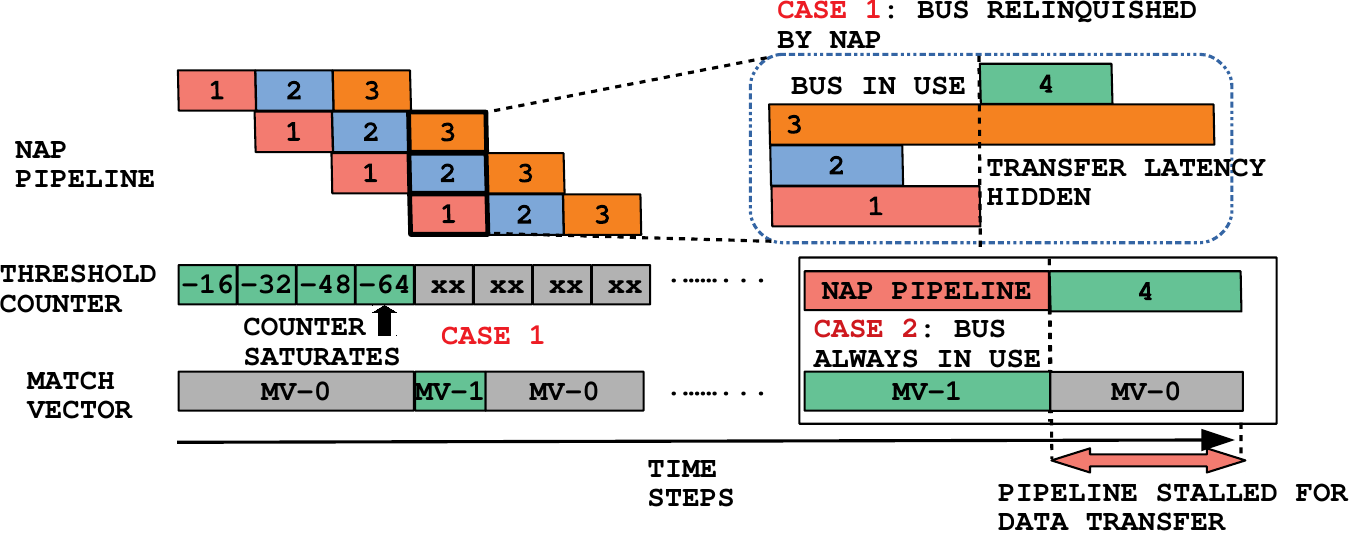, width = \columnwidth}
	\end{center}

	\caption{Different scenarios of data transfer pipelining within the DAM architecture
		\label{figure:pipeline_detailed}}

\end{figure}

\paragraph{Data Transfers} The transfer of data corresponding to a previous query can happen concurrently with the computation for the next query. However, Pipeline stages 1 and 2 utilize the data-bus and the data-arrays respectively, which requires stalling of such stages to enable data transfer. If the current pipeline stage within a given array is waiting for accumulation to complete, the group controller can transfer out data for the previous query through the data bus, thereby hiding the latency of such data transfer. If the accumulation stage fails to hide the complete data transfer latency, the pipeline is stalled till the data transfer completes. However, if stage 1 or 2 determines the execution time of a pipeline stage, it is not possible to transfer data during that stage. A worst-case situation is when none of the pipeline stages relinquish the data bus, which requires stalling of computation after current query to complete data transfer for the previous query. Both cases are illustrated in Figure~\ref{figure:pipeline_detailed}, using an example $I_{dist}$ in the range $48$ to $64$.

\subsection{Memory Organization}
Multiple memory chips are interfaced within a dual in-line memory module (DIMM) that offers scalability and parallelism across chips. A memory chip consists of multiple ranks, with each rank containing multiple chips. Each DIMM is interfaced with the CPU using the DDR protocol, resulting in a physical off-chip connection through 64 data pins. These 64 data pins are shared across all ranks, resulting in rank-level parallelism. Each chip within a rank is distributed across the 64 data-wires in an equal fashion. %For example, if each rank has 8 chips, each chip would have a dedicated 8-bit bus multiplexed across ranks.

\subsubsection{Memory Chip}
A memory chip consists of multiple memory groups interconnected using an H-tree to route commands($\mathbf{C}$)/address($\mathbf{A}$) to, and data($\mathbf{D}$) to/from each group to the chip periphery. A chip controller is implemented at the periphery of each chip to orchestrate operation across different groups. Each chip maintains a 64B register to store the incoming key from the CPU. Each successive key/mask entry is buffered at the periphery when its preceding key buffers are being executed within the memory, to hide the latency of off-chip key/mask transfers from the CPU. Once each memory group completes the required computation and data transfers corresponding to a single pipeline step, the chip controller issues a write to broadcast the next key buffer entry to all group controllers.

\paragraph{Tree-Based Index Generation} We add additional wires to the H-tree nodes between each group and the chip controller to support priority-encoding-based address generation for the candidate points, inspired by previous work~\cite{prasad2021memristive} to support indexing of data spread across groups. Each group controller maintains the smallest physical address of all valid candidate rows within its domain, internally generated through the group H-tree. This address is propagated upstream the H-tree, where each H-tree performs priority encoding of its leaves' addresses to extract the current active least address. Once its corresponding datapoint is read out and the corresponding match vector reset, this propagates upstream the H-tree and the index for the datapoints with the next-least physical address is generated. This approach allows the chip controller to buffer these generated addresses to perform bandwidth-efficient candidate fetches internally within the memory. These addresses corresponding to a single datapoint act as the candidate index.
For each datapoint transferred to the chip controller, it maintains a counter to keep track of the number of candidate points being fetched. If such a value increases beyond a parameterizable threshold, the chip controller insrtucts the groups to reset all threshold counters, signaling the end of search. To allow for sufficient time for H-tree propagation and candidate counters to update, the chip-controller instructs all the threshold counters to stall after every 32 cycles (determined through evaluation) to allow for interleaved data transfer and counter update.

\subsection{Area overhead Mitigation} Though each NAP contains only simple functional units, the overall design presents a non-trivial area overhead. A NAP unit per row within the memory array increases the total area of the memory by almost $6$x. To mitigate the same, we propose multiplexing of NAPs across $N_{MUX}$ consecutive rows within an array to mitigate the total area overhead. The group controller is given necessary additional wires as part of the H-tree to control the multiplexers across the different memory arrays to ensure correct operation under multiplexing of resources. However, as a consequence of the same, processing of all datapoints within each group now requires $N_{MUX}$ steps over all keys $k$, which increases overall latency and decreases overall throughput. A detailed analysis of the same is presented in the evaluation section.

\section{Software Support}

\paragraph{Configuration Parameters} The search quality of DAM is determined by two parameters - the bitmask $\mathbf{M}$ which defines the quantized range of values, and the coverage (or $\mathbf{Cov}$), which defines the degree of overfetch within every chip during data transfer. A decrease in bitmask $M$ results in an increase in quantized data-width, which increases compute latency by increasing both $NKeys$ and the number of bitwise searches required within the array at the benefit of higher precision. An increase in coverage results in an increase in the number of vectors to transfer accompanied by an increase in overall latency, but gives higher recall guarantees through re-ranking.

\paragraph{Address Mapping} The presence of multiplexing makes address mapping tricky, especially if each NAP can contain multiple vectors within its domain. All NAPs point to the same memory row across all groups to ensure data parallelism. This results in a logical partition of data that cuts across memory rows, arrays, groups, and chips, called a \textbf{Set}. Formally, we define a set as the collection of memory rows that maintain ownership of the NAP unit at the current iteration of computation. 
To keep computation latency at a minimum, it is important to map data to the memory on a "set-first" basis. Data spread across different sets cannot be computed upon in the same iteration. The memory controller uses the size of each vector datapoint to compute the number of rows required to store each datapoint. $NMux$ divided by this number defines the effective size of each memory set, which is then used during address mapping to interleave consecutive vectors across different sets.

\subsection{Programming model and API}
We design a userspace API library for DAM. This API enables users to 1) allocate memory in the accelerator 2) Set configuration parameters prior to each search 3) Generate $Key$ and $Dist$ data structures on the CPU and write them to all chips within the defined memory region and 4) fetch candidate points across all chips upon search completion.

Memory DIMMs are not capable of sending signals to the CPU without a prior request. Hence, the CPU polls to check for data availability across the DIMMs. Such polling is made efficient by leveraging information about the $Key$ dataset to accurately estimate the timing of data transfer availability. Once data is available at the chip periphery, the memory controller starts broadcasting requests across all sets that contain valid data. The chip controller receives each read request at the periphery, then performs a single data burst to send the 64-bit address of the datapoint out from the memory (required for candidate ID generation), which is followed by multiple data bursts to fetch the 512-bit cache line from the chip. Multiple chips participate in the data transfer by sending part of its candidate vector, resulting in peak bandwidth efficiency during data transfer. 

\section{Experimental Setup}
\label{section:setup}
%\subsection{Methodology}
\paragraph{Architecture} 
We model the operations of DAM with a lightweight memory simulator model based on ESESC~\cite{esesc} and interface the issuing of commands through a QEMU. To realize the API and software support, we modify QEMU for an extended version of memkind library~\cite{memkind} that enables special memory allocation required to initialize the search ranges and enables writes to configuration address space within the different DAM DIMMs. Table~\ref{table:parameters} shows DAM parameters.

\paragraph{Circuits}
\label{subsection:circuit}
The memory array and its associated sensing circuits, drivers, group controller, and interconnect elements are modeled using NVSim/NVMExplorer~\cite{dong2012nvsim, pentecost2021nvmexplorer} integrated with resistive memory parameters from prior work~\cite{wu2011low}.  All the additional gates, latches, and the control logic required for the near-array processing are synthesized using the Synopsys Design Compiler~\cite{DesignCo4} with GlobalFoundries~\cite{gfFinFETGlobalFoundries} at 12nm. All the SRAM units for the tables and data buffers at the chip controller, alongwith additional H-tree wiring are evaluated using CACTI 7.0~\cite{10.1145/3085572}. The match vectors and additional H-tree wiring incur a 5\% area overhead per group.

\paragraph{Baseline System}
We run the baselines on a 56-core Intel Xeon E5-2680 with a memory bandwidth of 102.4 GBps and measure execution time across the different kernels and workloads. Each datapoint is accumulated across multiple runs to average the noise in our baseline measurements.

\paragraph{Datasets}
We consider 4 popular billion-scale workloads (BIGANN, Microsoft-SPACEV, Facebook-SSNPP and Yandex-text2Image) representative of real workloads (embedding vectors) used in similarity search applications~\cite{simhadri2022results}, shown in Table~\ref{table:workloads}. We also evaluate million-scale splits for each of the datasets to measure performance at million-scale.
\begin{table}[h!]

	\caption{Workloads.\label{table:workloads}}
	\centering % Center the table

	\begin{tabular}{|p{1.2cm}|p{1cm}|p{1.5cm}|p{1.5cm}|p{1.5cm}|}
		\hline
		Name & Dims & Vectors & DataType & Distance \\
		\hline
		\textbf{BigANN} & 128 & 1B & uint8 & L2 \\
		\hline
		\textbf{MS-S} & 100 & 1B & int8 & L2 \\
		\hline
		\textbf{SSNPP} & 256 & 1B & uint8 & L2 \\
		\hline
		\textbf{TTI1B} & 200 & 1B & int8\footnotemark{} & IP \\
		\hline
	\end{tabular}
\end{table}

\footnotetext{Quantized from float32}

\paragraph{Algorithms}
We consider different compression and graph-based implementations as state-of-the-art baselines. % for similarity search applications. 
We consider \textit{IndexIVFPQ}~\cite{Structfa44} as the compression-based baseline and \textit{IndexHNSW}~\cite{Structfa14} as the graph-based baseline. The large indexing overheads of the graph approach make it prohibitive to faithfully model a billion-scale baseline using the given system configuration. For IVFPQ, we consider 2x and 4x compression for million-scale, and 8x and 16x compression for billion-scale data respectively, while varying
%to extract different baseline configurations based on 
the number of codewords ($k*$) and the number of subvectors ($M$). Different points within the recall/throughput vs latency space for each configuration are measured by varying $nprobe$ and $L$ for the different IVFPQ and HNSW baselines. $L$ defines the depth of greedy search in the HNSW graph. For DAM, we consider 3 bitmask values ($\mathbf{M}$ = $3$/$4$/$5$) each operating with three different coverage values ($\mathbf{Cov}$ = $1$/$5$/$10$).
We consider a tight metric of recall-$100@100$ for all our evaluations. 
\begin{table}[h!]
	\caption{Simulation parameters.\label{table:parameters}}
	\centerline
	{\scriptsize\setlength\tabcolsep{1.5pt}
		\begin{tabular}{|c|c|l|}
			\hline
			\multicolumn{2}{|c|}{\textbf{Core Type}}  & 56 Intel(R) Xeon(R) CPU E5-2680 v4 @ 2.40GHz\\
			\hline
			\hline
			\parbox[t]{2mm}{\multirow{3}{*}{\rotatebox[origin=c]{90}{\textbf{Cache}}}}
			& \textbf{Instruction L1} & 32KB, direct-mapped, 64B block \\
			\cline{2-3}
			& \textbf{Data L1}        & 32KB, 4-way, LRU, 64B block \\
			\cline{2-3}
			& \textbf{Data L2}      & 256KB, 16-way, LRU, 64B block \\
			\cline{2-3}
			& \textbf{Data L3}      & 32MB, 32-way, LRU, 64B block \\
			\hline
			\hline
			\parbox[t]{2mm}{\multirow{4}{*}{\rotatebox[origin=c]{90}{\textbf{Main}}}}
			& \textbf{Memory} &        256GB, 4KB row buffer, DDR4-1600\\
			& \textbf{Configuration}  & Channels/DIMMs/Ranks/Chips: 8/1/2/16. 8 Gb chips\\
			\cline{2-3}
			& \textbf{Timing}         & tRCD:44, tCAS:44, tCCD:16, tWTR:31, tWR:4, tRTP:46, tBL:10\\
			& \textbf{(CPU cycles)}  & tCWD:61, tRP:44, tRRD:16, tRAS:112, tRC:271, tFAW:181\\
			\hline
			\hline
			\parbox[t]{2mm}{\multirow{5}{*}{\rotatebox[origin=c]{90}{\textbf{DAM}}}}
			& \textbf{Memory}  & 256GB, 4KB row buffer, DDR4-1600, 16 8Gb DAM chips/rank. \\
			& \textbf{Configuration} &  Channels/DIMMs/Ranks/Chips: 8/1/2/32. \\
            & \textbf{  } & 256 512x512 SLC subarrays/group. 128 groups/chip \\
			\cline{2-3}
			& \textbf{NAP/Memory} & $t_{\mathrm{bit}}$:1, $t_{\mathrm{writebuf}}$:64, $t_{\mathrm{count}}$:1, $t_{\mathrm{add}}$:8\\
                & \textbf{Parameters} & $T_{\mathrm{read}}$:4.3ns, $T_{\mathrm{write}}$:54.2ns, $v_{\mathrm{Write}}$:2V, $v_{\mathrm{Compute}}$:1V\\
			\hline
		\end{tabular}
	}
\end{table}

\paragraph{Emerging Applications}
The Yandex-text2Image dataset works with cross-modal queries (text-based embedding queries on image-based embedding), where the queries are out-of-distribution in comparison to the dataset. Index freshness is a trivial problem for our proposed approach, which just needs the corresponding memory range to be initiated and the datapoints to be written into the memory range. Filtered queries can be incorporated during $Key$ datastructure generation through masking of the corresponding dimensions that need to be excluded from search.

\section{Evaluations}
\label{experiments}
\subsection{Area Analysis}
\begin{figure}[h]
	\begin{center}

		\epsfig{file=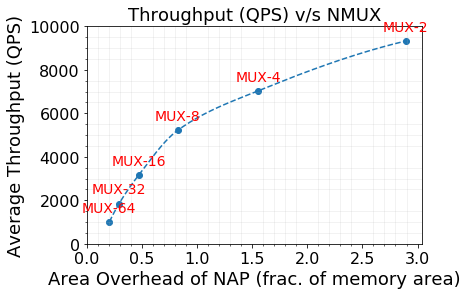, width = 0.7\columnwidth}
	\end{center}

	\caption{Area overhead v/s Throughput analysis to identify optimal multiplexing
		\label{figure:area/execbreakdown}}
\end{figure}

% \begin{figure*}[h]
% 	\vspace{-2ex}
% 	\begin{center}
% 		\epsfig{file=figures/latency.png, width = \linewidth}
% 	\end{center}
% 	\vspace{-3ex}
% 	\caption{Latency comparison for DAM and different software implementations
% 		\label{figure:latency}}
% 	\vspace{-2ex}
% \end{figure*}
To identify the optimal multiplexing configuration for DAM, we plot the average throughput of a DAM chip across all datasets using all chosen bitmask and coverage parameters in Figure~\ref{figure:area/execbreakdown}. The X-axis shows the area overhead of all additional circuitry within each memory chip as a fraction of the chip area, while the Y-axis shows the average throughput across all measured configurations.
%within the chip. 
$N_{Mux}$ corresponds to the number of rows multiplexed with one NAP unit within an array. We observe that throughput steadily increases with the increasing area overhead, which tapers off significantly between $N_{Mux}=8$ and $N_{Mux}=16$. This throughput stagnation is owing to the shift of bottleneck from compute to external bandwidth with the decreasing value of $N_{Mux}$.
To account for peak power limitations, we proceed with a conservative choice of $N_{Mux} = 16$ for our 
%million and billion-scale 
evaluations. 

%To identify the impact of varying $N_{Mux}$ on the latency for each query, we plot a breakdown of the average time spent in compute v/s latency across all configurations and Datasets in figure~\ref{figure:area/execbreakdown}. The total latency is significantly dominated by data transfer in lower $N_{mux}$ values. However, with increasing $N_{Mux}$, the computation time starts dominating with more opportunity for hiding data transfer latency, reflected in the graph. We proceed with a conservative choice of $N_{Mux}=16$ for the subsequent evaluations.

\subsection{Million-Scale Evaluations}
\begin{figure*}[h]
	\begin{center}
		\epsfig{file=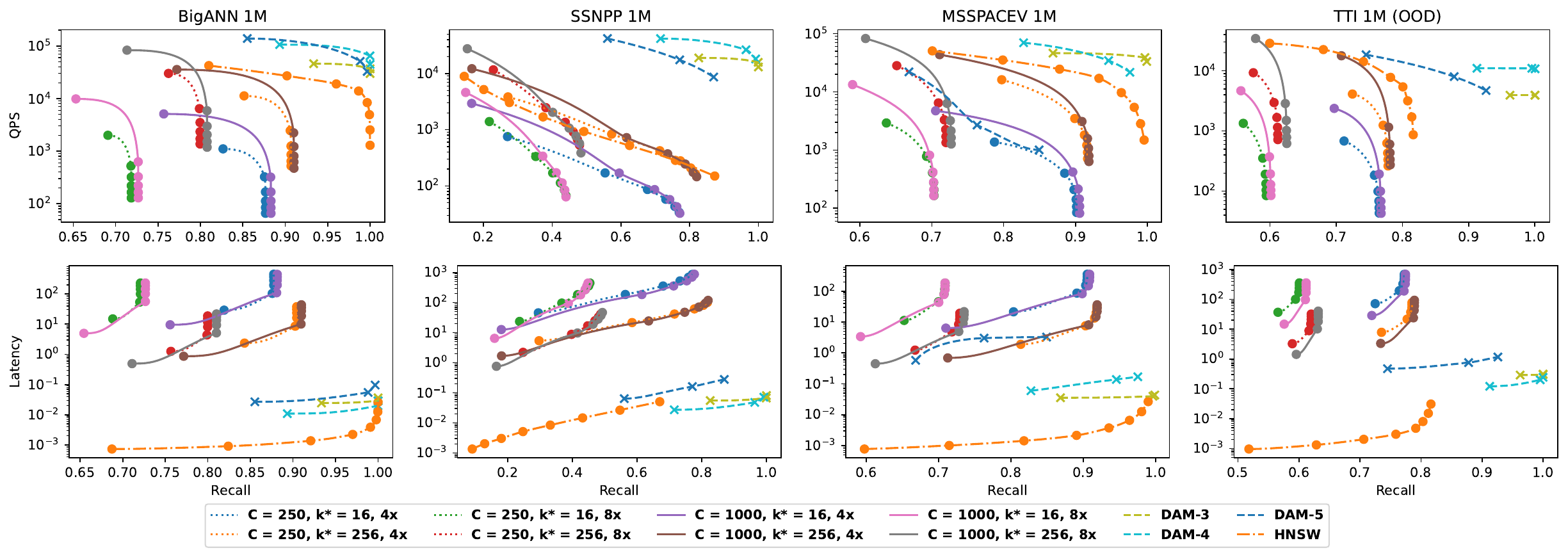, width = 0.95\linewidth}
	\end{center}

	\caption{Throughput and Tail Latency comparison on different 1M dataset-splits between DAM and baselines.
		\label{figure:1M}}
\end{figure*}

Figure~\ref{figure:1M} shows the throughput and latency improvements across different configurations of the IVFPQ and HNSW baselines on 1M data splits. Each PQ trend is characterized by three values - $|C|$, $k*$, and \textit{Compression Ratio}. Each trend captures the behavior of the underlying index under varying degrees of search-space pruning, controlled using $nprobe$ and $L$\footnotemark{} in IVFPQ and HNSW respectively. The DAM configuration points are denoted by an "x", while the baselines are denoted by dots.
\footnotetext{$L$ defines the depth of the search in the HNSW graph}

\paragraph{Throughput Analysis} The top half of Figure~\ref{figure:1M} shows the queries-per-second (QPS) v/s recall trade-off space for the different baselines and DAM. The top-right corner represents ideal performance. Most DAM configurations are situated near the most optimal corner of the  QPS v/s recall space. Each DAM trend line corresponds to a specific $\mathbf{M}$, with trends shown for varying $\mathbf{Cov}$. We see that the DAM and HNSW baselines can achieve high recall while operating at a better throughput wrt the PQ throughputs, owing to support for re-ranking. Majority of this QPS improvement can be attributed to the proposed address mapping optimized to minimize the number of iterations, on top of the reduction in bandwidth-requirement. In MSSPACEV, we notice that an increase in $\mathbf{M}$ from $4$ to $5$ results in a significant QPS degradation. This is attributed to the aliasing effects of approximate distance calculated within the quantized space, with a large fraction of vectors having the same quantized distance. This effect is exacerbated by an increase in coverage. Analysis on the TTI 1M dataset shows that DAM can achieve almost perfect recall even with out-of-distribution queries, while HNSW plateaus at around $0.8$ recall. The negligible increase in QPS DAM-3 and DAM-4 in TTI can be attributed to the nature of Inner-Product computation within DAM being bound by the NAP accumulation stages. This allows for significant latency hiding through pipelined data transfers. 
%As expected, an increase in $\mathbf{Cov}$ is accompanied by a corresponding increase in recall, while an increase in $\mathbf{M}$ results in a decrease in recall
\paragraph{Latency Analysis} The bottom half of Figure~\ref{figure:1M} shows the 99th percentile tail latencies for the different configurations. HNSW has better tail latencies at lower recall spaces owing to the small working-set size, resulting in decreased average memory access time. However, with increasing $L$, the latencies converge towards DAM, denoting that the working-set size exceeds the processor cache size.
%inability of the processor cache to capture the corresponding increase in working-set size effectively. 
The aliasing effects are distinctly visible for DAM-5 MSSPACEV, with a drastic increase in latency in comparison to DAM-4. The increase in $\mathbf{Cov}$ only results in a minimal increase in tail latency across most configurations. This is due to DAM's bandwidth-efficient transfer of data.
DAM achieves 8.2x/100x/9x/2x and 972x/985x/85x/100x throughput and latency improvement over the best PQ baseline configuration across the 4 datasets respectively.

\begin{figure*}[h]
	\begin{center}
		\epsfig{file=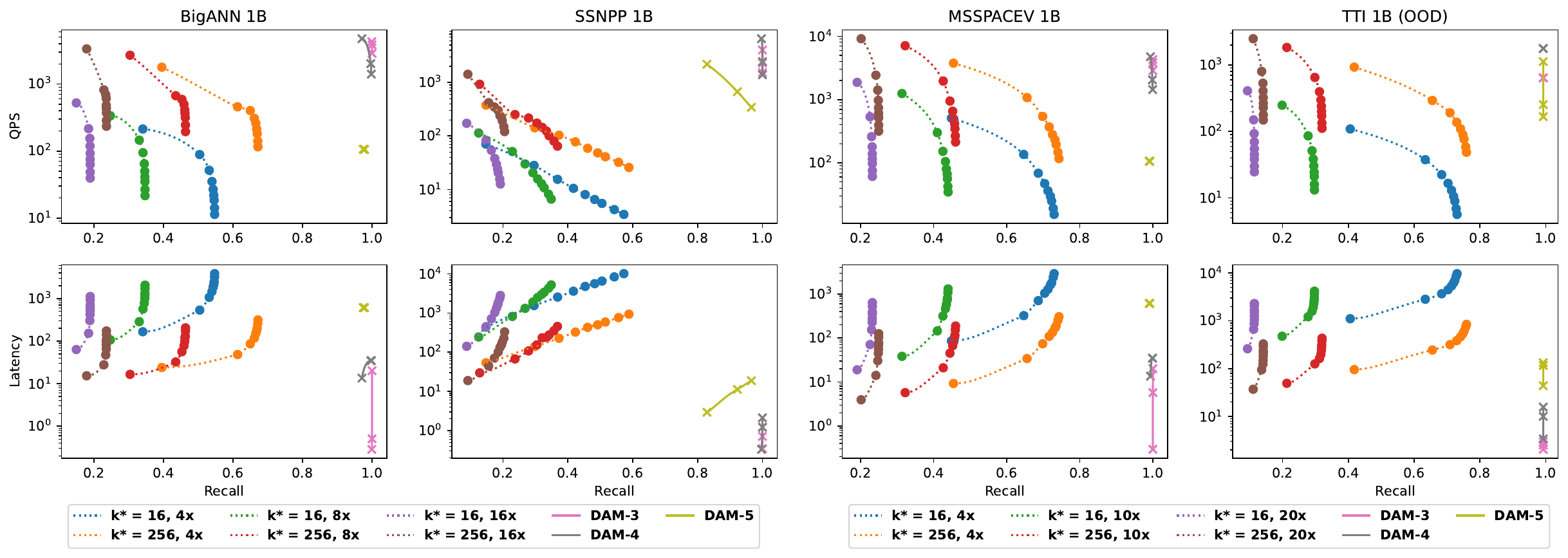, width = 0.95\linewidth}
	\end{center}

	\caption{Throughput and Tail Latency comparison on 1B indices between DAM and baselines.
		\label{figure:1B}}
\end{figure*}
\subsection{Billion-Scale Evaluations}
Figure~\ref{figure:1B} shows the throughput and latency improvements across different configurations of the IVFPQ and HNSW baselines on the 1B datasets. Index construction was performed on chunks of 500 million owing to memory limitations. We run each chunk independently with full available bandwidth and take a max of the two execution times to measure the final execution time. We perform data re-ranking across the  obtained candidates from both to generate the final recall. Such a setup acts as an %\textcolor{red}{definite} 
upper bound for true baseline both for recall and latency.

\paragraph{Throughput analysis}
The baseline QPS trendlines follow the same pattern as the 1M case but saturate at a much lower recall. This is due to the inability of the Product Quantizer to capture the data distribution accurately with increasing data scale. DAM achieves very high recall irrespective of the dataset across all values of $\mathbf{M}$ and $\mathbf{Cov}$. This is a result of the inherent overfetch of data enabled by chip-level parallelism. However, the fraction of the entire dataset visited per query still remains a low percentage ($\le 10^{-3}\%$).  Moreover, such data transfer happens in a highly bandwidth-efficient fashion. There is a sharp decrease in QPS with the increase in coverage for SSNPP and TTI with DAM-5, due to the aliasing effect at lower precisions. Interestingly, MSSPACEV does not suffer this problem at billion scale, owing to the abysmal performance of the DAM-3 $Cov = 1$, which obfuscates any further change in throughput with increasing coverage.

%For example, the SSNPP dataset is spread across 128 memory chips, and each chip needs to return at least $k$ (Nearest-Neighbor) values to ensure high recall. The framework needs to fetch $12800$ values at the minimum for every single query, which in relative terms is just $1.28 \times 10^{-3}\%$ of the entire dataset.
\paragraph{Latency analysis}
Unlike the 1M case, there is an inverse correlation between the trends observed in the baseline throughput and recall. This shows that PQ is severely bandwidth-bound at large data scales. BIGANN and MSSPACEV suffer from increased effects of coverage on tail latency due to the low latencies achieved by $\mathbf{M}=1$. On the other hand, increasing coverage of DAM-5 hardly results in a latency increase owing to the poor latency achieved by $\mathbf{M}=1$.
DAM achieves 13.3x/76.2x/4.8x/1.3x and 937x/603x/3.8x/1.9x throughput and latency improvement over the best PQ baseline configuration across the 4 datasets respectively, with DAM achieveing near-perfect recall across all configurations.

\subsection{Energy and Power Analysis}
\paragraph{Power} For the selected configuration of $N_{Mux}=16$, each group controller with its associated NAP, wiring, and buffers operates at a peak power of $382uW$. The peak power of each proposed 8Gb memory chip is around 0.488W. Most of this power consumption occurs from the switching activity and data movement associated with the NAP, the group, and chip controllers to enable the required computation. In practice, the actual power is lower because of the intermittent pipeline stalls incurred during data transfer. 

\paragraph{Energy} Though a DAM chip operates at a higher power limit than a conventional memory DIMM, the proposed memory system achieves $28.5\times$, $8.57\times$, $67.4\times$, $2.21\times$ energy improvements in comparison to the most optimal configuration of BIGANN, MS-SPACEV, SSNPP, and TTI respectively. This is predominantly due to the massive reduction in data movement enabled by DAM.

\section{Related Work}
\label{section:related}

\paragraph{Similarity Search Algorithms}
There are many software approaches to ANNS,
%There exist several software approaches to ANNS excluding HNSW and Product Quantization, which follow the trend of being compression-based, graph-based, or hybrid. A comprehensive survey exists on the different prevalent 
including graph-based~\cite{wang2021comprehensive,jaromczyk1992relative, GitHubmi14, fu2017fast,subramanya2019diskann}.
%}, with various implementations based on the relative neighbor graph~\cite{} most popular excluding HNSW being \textit{Vamana}~\cite{subramanya2019diskann}.
DiskANN~\cite{GitHubmi65} is a suite of different \textit{Vamana}~\cite{subramanya2019diskann} based graph algorithms, with support for filtered queries~\cite{10.1145/3543507.3583552} and index freshness~\cite{singh2021freshdiskann}. However, such graph-based indices incur significant memory overhead to store metadata and are hence scaled out to disks. The cost model in such cases differs owing to the primary bottleneck being SSD-round trip access latency~\cite{subramanya2019diskann}. We instead focus on the in-memory acceleration of similarity through a novel algorithm design to fit the proposed distance-based addressable memory model. 
Compression-based approaches are typically based on product quantization, e.g., OPQ~\cite{ge2013optimized} and Google ScaNN~\cite{guo2020accelerating}. %OPQ~\cite{ge2013optimized} applies rotation to the underlying cartesian space to optimize distribution for search. 
%Google ScaNN~\cite{guo2020accelerating} utilizes product quantization inspired by a novel anisotropic loss function. 
However, PQ fundamentally lacks re-ranking support.
%the fundamental issue remains of PQ still remains the lack of re-ranking support.
Alternative techniques employ locality-sensitive hashing (LSH)~\cite{zheng2020pm, shrivastava2014asymmetric, GitHubFA43, tian2023db} or trees~\cite{GitHubsp12}, but such approaches perform poorly at scale. Several startups offer ANNS-as-a-service~\cite{10.1145/3448016.3457550, VectorDa30, Vespa} with support for many of the above algorithms.

\paragraph{Hardware Acceleration of ANNS}
ANNA~\cite{ANNA} proposes an accelerator for PQ through memo-ization 
%and shows significant performance improvement over CPU and GPU baselines under 
with relaxed recall requirements. Such an approach fails to scale to high-recall at tighter recall guarantees owing to inherent algorithmic limitations. TPU-KNN~\cite{chern2022tpu} leverages Google TPUs~\cite{46078} through a novel algorithm to accelerate ANNS. FPGA-based accelerators~\cite{8977838, zhang2018efficient} for PQ fail to scale to billion-scale data. Tigris~\cite{xu2019tigris} proposes a similarity search accelerator, but exclusively in the context of 3D pointcloud processing.

\paragraph{Processing-In-Memory}
Resistive RAM-based processing-in-memory has been 
%a significant topic of research over the past few years, ranging from crossbar accelerators 
deployed for machine learning~\cite{shafiee2016isaac, imani2019floatpim, yuan2021forms} to more general-purpose architectures for bitwise processing~\cite{truong2021racer, kvatinsky2014magic} for data-parallel applications~\cite{perach2023understanding}. However, many of these works
%such work relies on processing within the memory arrays, thereby incurring 
require a significant number of memory writes, which limits the lifetime of the cells. We rely on a bit-serial processing model that does not require any writes to the memory during computation.

\section{Conclusions}
\label{conclusions}

High-dimensional ANNS is an important emerging problem due to the prevalence of vector databases. However, most existing approaches fail to achieve high recall as dataset scale increases, because they employ pruning mechanisms to circumvent the bandwidth bound. This paper proposes a novel memory system ("DAM") capable of accessing data based on its approximate proximity to a given query. By employing pruning as a post-processing step within memory, DAM is able to achieve high recall irrespective of dataset characteristics, while offering significant performance and energy improvements over state-of-the-art software baselines.  

\section{Acknowledgement}

This work was supported in part by NSF grant CCF-2217154.

\bibliographystyle{plain}
\bibliography{refs.bib}

\end{document}